# Ratio of Mediator Probability Weighting for Estimating Natural Direct and Indirect Effects

Guanglei Hong
University of Chicago, 5736 S. Woodlawn Ave., Chicago, IL 60637

**Abstract**
Decomposing a total causal effect into natural direct and indirect effects is central to revealing causal mechanisms. Conventional methods achieve the decomposition by specifying an outcome model as a linear function of the treatment, the mediator, and the observed covariates under identification assumptions including the assumption of no interaction between treatment and mediator. Recent statistical advances relax this assumption typically within the linear or nonlinear regression framework with few exceptions. I propose a non-parametric approach that also relaxes the assumption of no treatment-mediator interaction while avoiding the problems of outcome model specification that become particularly acute in the presence of a large number of covariates. The key idea is to estimate the marginal mean of each counterfactual outcome by weighting every experimental unit such that the weighted distribution of the mediator under the experimental condition approximates their counterfactual mediator distribution under the control condition. The weight to be applied for this purpose is a ratio of the conditional probability of a mediator value under the control condition to the conditional probability of the same mediator value under the experimental condition. A non-parametric approach to estimating the weight on the basis of propensity score stratification promises to increase the robustness of the direct and indirect effect estimates. The outcome is modeled as a function of the direct and indirect effects with minimal model-based assumptions. In contrast with the regression-based approaches, this new method applies regardless of the distribution of the outcome or the functional relationship between the outcome and the mediator, and is suitable for handling a large number of pretreatment covariates. Under modified identification assumptions, the weighting method also makes adjustment for post-treatment covariates, a benefit apparently unavailable in the existing methods.

**Key Words**: Causal inference; Counterfactual outcomes; Identification; Mediation; Non-parametric method; Post-treatment covariates; Treatment-mediator interaction.

## 1. Introduction

The goal of many scientific investigations is to not just examine whether input $A$ would generate output $Y$ but also discern among competing theories explaining why $A$ causes $Y$. If $A$ changes $Z$ which subsequently changes $Y$, $Z$ is considered to be a mediator. The remaining effect of $A$ on $Y$ that has not been channeled through $Z$ has been called the direct effect. For example, we might ask whether attending a federally funded Head Start program improves the attention skills of children from low-income families and thereby reduces their likelihood of repeating kindergarten. A major advancement in the recent statistical literature on mediation has been the clarification of conceptual distinctions between controlled direct effects and natural direct effects (Pearl, 2001, Robins & Greenland, 1992). A controlled direct effect of attending a Head Start program on reducing the likelihood of repeating kindergarten retention is conceivable if all low-income children would display the same level of attention at kindergarten entry as a result of a second intervention regardless of their prior preschool experience. In contrast, a natural direct effect represents the change in retention status attributable to attending a Head Start program when the child's attention skills at kindergarten entry would have counterfactually remained unchanged by the Head Start attendance. In many situations, scientific questions about natural direct and indirect effects are especially relevant for understanding causal mechanisms.

In general, if the controlled direct effect of the treatment on the outcome does not depend on any mediator value, under identification assumptions explicated by previous researchers (Holland, 1988; Pearl, 2001; Robins & Greenland, 1992; Sobel, 2008), the natural direct effect and the controlled direct effect will become equal. Researchers can rely on path analysis, structural equation modeling, or similar regression models to obtain estimates of natural direct and indirect effects by invoking model-based assumptions. However, in the cases where the treatment and the mediator have an interaction effect on the outcome, the standard methods no longer apply. A number of alternative methods have been proposed recently for estimating the natural direct effect and the natural indirect effect. They typically involve specifying an outcome model as a function of the treatment, the mediator, and the observed covariates (Pearl, 2010; Petersen et al, 2006; VanderWeele, 2009). In addition, these methods require model-based assumptions about the association between the outcome and the mediator. Imai, Keele, and Yamamoto (2010) have developed a non-parametric procedure for estimating the natural indirect effect. Robins (2003) argued in the presence of post-treatment covariates, estimation of natural direct and indirect effects requires the assumption of no treatment-by-mediator interaction. Hence none of the existing methods are capable of handling post-treatment covariates that confound the mediator-outcome relationship when there is a treatment-by-mediator interaction.

I propose an alternative non-parametric approach that allows for a simultaneous estimation of the natural direct and indirect effects regardless of whether there is a treatment-by-mediator interaction effect on the outcome. I use weighting to not only adjust for selection bias associated with observed confounders but also approximate the counterfactual mediator distribution associated with an alternative treatment condition. The weighted outcome is modeled only as a function of the direct and indirect effects of interest. Hence, it does not require any parametric specification of the functional relationships between the outcome and the treatment, between the outcome and the mediator, or between the outcome and the observed covariates. Because the outcome model is non-parametric, this analytic approach is applicable regardless of the distribution of the outcome, the distribution of the mediator, or the functional relationship between the two. I show that in the absence of post-treatment covariates, the weight can be derived easily under standard identification assumptions. In the presence of post-treatment covariates, a modification of the standard assumptions and an addition of another assumption enable us to derive the weight that adjusts for selection bias associated with both pretreatment and post-treatment covariates.

I organize this paper as follows. Section 2 introduces notation in the context of an illustrative application in which natural direct and indirect effects are of particular interest. Section 3 reviews the existing standard and alternative approaches to estimating the natural effects. Section 4 lays out the theoretical rationale for the new approach to estimating natural direct and indirect effects. This section also explains the analytic procedure in the context of the application example. Section 5 summarizes the major features of the new approach and raises issues for further inquiry.

## 2. Definition of Natural Direct and Indirect Effects: An Illustrative Application

### 2·1 Notation

To illustrate in a simple setting, let $A = 1$ if a child has attended a Head Start program and let $A = 0$ represent the experience of growing up in a non-Head Start setting during the preschool years. Let us suppose that all schools use the same screening criteria to determine whether a new kindergartner is displaying attention skills at either a low level or a high level denoted by $Z = 0$ and $Z = 1$, respectively. The outcome is denoted by $Y$, with $Y = 1$ indicating that a child is retained in kindergarten and $Y = 0$ indicating that the child is not retained. Adopting Rubin's Causal Model (Holland, 1986, 1988; Rubin, 1978) and invoking the stable unit treatment value assumption (SUTVA) for simplicity (Rubin, 1986), I use $Z_a$ to denote a child's potential mediator and $Y_{aZ_a}$ for the child's potential final outcome. We would observe a child's attention skills at the level $Z_1$ if the child has attended a Head Start program and $Z_0$ otherwise. If the child has obtained a high attention level after attending a Head Start program, the child would show retention status $Y_{11}$; if the child has obtained a low attention level after attending a Head Start program, the child would show retention status $Y_{10}$ instead. If the child has grown up in a non-Head Start setting and nonetheless has obtained a high attention level, the child would show retention status $Y_{01}$; otherwise, if the child has grown up in a non-Head Start setting and has obtained a low attention level, the child would show retention status $Y_{00}$. Additional counterfactual outcomes are denoted by $Y_{aZ_{a'}}$. For example, $Y_{1Z_0}$ represents the child's retention status associated with attending a Head Start program yet the child's attention skills were set at the level that the child would have displayed without attending Head Start.

### 2·2 Controlled Direct Effects

Following Pearl's (2001) framework, the controlled direct effect is defined by $Y_{1z} - Y_{0z}, z \in (0,1)$ for a given child. However, it is possible that in a subset of the population, the controlled level $z$ is different from these children's potential mediator values $Z_1$ and $Z_0$. Hence the question about the controlled direct effects has low practical relevance in this case.

### 2·3 Natural Direct Effects and Natural Indirect Effects

The natural direct effect is defined by $Y_{1Z_0} - Y_{0Z_0}$ for a given child. Here the question becomes whether a Head Start program can reduce the likelihood of grade retention without raising low-income children's attention skills. If assigned to a non-Head Start setting, children's attention level $Z_0$ at kindergarten entry varies naturally in the population. Hence, to take the expectation of the natural direct effect requires a summation of the controlled direct effects over the distribution of $Z_0$, that is, $E(Y_{1Z_0} - Y_{0Z_0}) = \sum_z E(Y_{1z} - Y_{0z}) pr(Z_0 = z)$ (Pearl, 2001). The natural indirect effect, defined by $Y_{1Z_1} - Y_{1Z_0}$ for a given child, is the change in retention status caused only by the Head Start-induced change in the child's attention skills from $Z_0$ to $Z_1$ at kindergarten entry. The total effect of attending a Head Start program vs. a non-Head Start setting on retention status is the sum of the natural indirect effect and the natural direct effect: $Y_{1Z_1} - Y_{0Z_0} = (Y_{1Z_1} - Y_{1Z_0}) + (Y_{1Z_0} - Y_{0Z_0})$. I will focus my discussion on how to analyze the natural direct and indirect effects as defined above.

### 2·4 Pretreatment and Post-Treatment Covariates

Let $X$ denote a set of pretreatment covariates that may confound the treatment-mediator relationship, the treatment-outcome relationship, or the mediator-outcome relationship. For example, minority status is a pretreatment covariate that may influence a child's participation in a Head Start program, the child's attention level at kindergarten entry, as well as the child's likelihood of repeating kindergarten. Let $L_a$ denote a set of post-treatment covariates that may result from having been assigned to treatment $a$ and may subsequently confound the mediator-outcome relationship. For example, Head Start may improve children's physical well-being which may in turn raise their attention level and may also prevent school absence due to health problems in kindergarten and thereby reduce the likelihood of grade retention.

### 3. Existing Approaches to Estimating Natural Direct and Indirect Effects

### 3·1 Path Analysis: the Standard Approach

In this section I review the assumptions under which currently available methods generate consistent estimates of the natural direct and indirect effects defined above. Social scientists have employed path analysis as primary tools for identifying mediators of treatment effects on final outcomes (Baron & Kenny, 1986; Duncan, 1966). The analysis typically involves two linear regression models for a continuous outcome $Y$ and a continuous mediator $Z$ each specified as a linear function of treatment assignment $A$:
$$Z = \alpha_Z + dA + \epsilon_Z, \qquad Y = \alpha_Y + bZ + cA + \epsilon_Y,$$
where $\epsilon_Z$ and $\epsilon_Y$ are structural errors. Researchers usually interpret $c$ as the direct effect of $A$ on $Y$ and interpret the product $bd$ as the indirect effect of $A$ on $Y$ mediated by $Z$. In this framework, the total effect of $A$ on $Y$, which can be obtained by regressing $Y$ on $A$, is equal to $bd + c$.

Using the potential outcomes framework, researchers (Holland 1988; Imai et al, 2010; Pearl, 2001; Robins and Greenland, 1992; Sobel, 2008; VanderWeele & Vansteelandt, 2009) have clarified the identification assumptions under which the above path coefficients have causal meanings. Suppose that pretreatment covariates $X$ can possibly be controlled for through either research design or statistical adjustment. The assumptions include, for all values of $a, a', z$, and $z'$:

Assumption 1 (Nonzero probability of treatment assignment). $0 < pr(A = a \mid X) < 1$.

Assumption 2 (Nonzero probability of mediator value assignment within a treatment). $0 < pr(Z_a = z \mid A, X) < 1$.

Assumption 3 (No confounding of treatment-outcome relationship). $Y_{az} \amalg A \mid X$.

Assumption 4 (No confounding of mediator-outcome relationship within the actual treatment condition). $Y_{az} \amalg Z_a \mid A = a, X$.

Assumption 5 (No treatment-by-mediator interaction). $E(Y_{az} - Y_{a'z}|X) = E(Y_{az'} - Y_{a'z'}|X)$.

When the functional form of the linear structural model is correctly specified, the above assumptions suffice for evaluating the controlled direct effect represented by path coefficient $c$.

Even when all the above assumptions hold, to estimate natural direct and indirect effects with the standard methods requires additional identification assumptions (Pearl, 2001):

Assumption 6 (No confounding of treatment-mediator relationship). $Z_a \coprod A \mid X$.

Assumption 7 (No confounding of mediator-outcome relationship across treatment conditions). $Y_{az} \coprod Z_{a'} \mid A = a, X$.

Under Assumptions 1--7, $c$ represents the natural direct effect, and $bd$ represents the natural indirect effect.

The standard methods have some major limitations. When there is a treatment-by-mediator interaction, the controlled direct effect will depend on the fixed level of the mediator. Path coefficient $c$ no longer represents the natural direct effect except when $d$ is zero, that is, when $Z$ does not qualify as a mediator between $A$ and $Y$. Path analysis can make covariance adjustment for a rather limited number of pretreatment covariates and is prone to misspecifications of the functional form of the outcome model and the mediator model. In addition, when post-treatment covariates confound the mediator-outcome relationship, the standard methods through regression cannot adjust for such covariates without biasing the estimate of the causal effect of $A$ on $Z$ and that of the direct effect of $A$ on $Y$ (Rosenbaum, 1984).

### 3·2 Alternative Methods for Estimating Natural Direct and Indirect Effects

The assumption that treatment and mediator do not interact in causing the outcome is often implausible. Statisticians have recently proposed alternative approaches that relax the no-interaction assumption in estimating natural direct and indirect effects. These include the modified regression approach (Petersen et al, 2006) and the conditional structural models approach (VanderWeele, 2009). These new approaches represent an important advance in the research methodology for mediation.

Petersen, et al (2006) modified the regression approach to estimating natural direct effects by directly incorporating an interaction between treatment and mediator in the outcome model. This approach still requires Assumptions 1 to 4 and 6. However, they replaced assumptions 5 and 7 with "the direct effect assumption", that is, the controlled direct effect at the fixed value $z$ does not depend on $Z_0$, $E(Y_{az} - Y_{0z}|A,X) = E(Y_{az} - Y_{0z}|Z_0 = z, A, X)$. For the direct effect assumption to hold, it is usually necessary to identify pretreatment characteristics that predict both $Z_0$ and $Y_{az} - Y_{0z}$. The above assumptions suffice only for estimating the natural direct effect. To estimate the natural indirect effect, one must subtract the estimated natural direct effect from the estimated total effect. The estimation of the latter requires a different assumption that is generally stronger than Assumption 3.

Assumption 8. (Independence of treatment assignment and potential outcomes). $Y_{aZ_a}, Y_{a'Z_a}, Y_{aZ_{a'}}, Y_{a'Z_a} \amalg A \mid X$.

This assumption usually holds when treatment assignment is randomized or approximates a randomized experiment within levels of the pretreatment covariates.

Petersen et al's analytic procedure involves three major steps. Step 1 is to analyze a multiple regression of the outcome on the mediator, treatment, pretreatment covariates, and their interactions $E(Y \mid A, X, Z)$ and obtain the corresponding regression coefficient estimates, from which one can compute the controlled direct effect when $Z = z$ as a function of $X$ and test the null hypothesis of no direct effect. Step 2 is to analyze a multiple regression of the mediator on the treatment and pretreatment covariates $E(Z \mid A, X)$, from which one obtains an estimate of $E(Z_0 \mid X)$ when $A = 0$ and then compute the sample estimate of $E(X)$, $E(Z_0)$, and $E(Z_0 X)$. Step 3 is to compute the average direct effect of the treatment on the outcome $\hat{E}(Y_{1Z_0} - Y_{0Z_0}) = \beta_1 + \beta_2 \hat{E}(X) + \beta_3 \hat{E}(Z_0) + \beta_4 \hat{E}(Z_0 X)$. Similar to the standard approach of path analysis, the modified regression approach requires linearity of the model for $Y$ and of the model for $Z$. The regression models become cumbersome when there are a relatively large number of pretreatment covariates.

To estimate both natural direct effects and natural indirect effects, VanderWeele (2009) employed two conditional structural models similar to the regression models specified in Petersen et al's (2006) steps 1 and 2. The conditional structural models approach invokes Assumptions 1, 2, 4, 6, 7, and 8. To relax Assumption 5, this approach again includes an interaction between the treatment and mediator in the outcome model. Yet several major differences distinguish these two approaches. Instead of using the direct effect assumption, the conditional structural models are made conditional on a subset of pretreatment covariates that are required for meeting Assumption 7. Let $\Omega$ denote the entire set of pretreatment covariates and let $X_{(\text{vii})} \in \Omega$ represent the subset of pretreatment covariates required for meeting Assumption 7. The conditional structural models are $E(Y_{az} \mid X_{(\text{vii})} = x_{(\text{vii})}) = g(a, z, x_{(\text{vii})})$ for the effects of the treatment and mediator on the outcome and $E(Z_a \mid X_{(\text{vii})} = x_{(\text{vii})}) = h(a, x_{(\text{vii})})$ for the effect of the treatment on the mediator. The confounding effects of all pretreatment covariates $X$ in $\Omega$ are adjusted through inverse-probability of-treatment weighting. When the above mentioned identification assumptions as well as the model-based assumptions hold, a weighted analysis of the two conditional structural models generates consistent estimates of the coefficients. Further assuming that $g(a, z, x_{(\text{vii})})$ is a linear function of $z$, one can obtain an estimate of the counterfactual outcome $E\left(Y_{aZ_{a'}} \mid X_{(\text{vii})} = x_{(\text{vii})}\right) = g(a, h(a', x_{(\text{vii})}), x_{(\text{vii})})$. The result is consistent with that derived by Petersen et al (2006).

A potential advantage of the conditional structural models approach, when compared with the modified regression approach, is that the inverse-probability-of-treatment weighting strategy enables adjustment for a relatively large number of pretreatment covariates. However, as the subset of covariates related to Assumption 7 increases, the above advantage vanishes and the conditional structural models become cumbersome. The conditional structural models approach and the modified regression approach share a number of other limitations. Both strategies require combining parameters estimated from the outcome model and the mediator model to compute the point estimates of natural direct and indirect effects without simultaneous estimation of the confidence intervals; both approaches are prone to specification errors in the functional form of the outcome model and the mediator model; both procedures apply only to continuous outcomes that have a linear relationship with the mediator; and finally, in the presence of treatment-mediator interactions, neither procedure is suitable if a post-treatment covariate confounds the mediator-outcome relationship.

Viewing the counterfactual outcomes as missing data, van der Lann and Petersen (2005, 2008) outlined a series of methods for estimating the natural direct effect. These included inverse-probability-of-censoring-weighted estimation method, double robust inverse-probability-of-censoring-weighted estimation method, and likelihood regression-based estimation method. All these methods require a user-supplied parameterization of the natural direct effect and additional modeling assumptions to obtain estimators with good practical performance.

Imai and his colleagues (2010) developed a non-parametric estimation procedure under the assumption of sequential ignorability. In essence, the treatment is assumed to be ignorable given the pretreatment covariates; and the mediator is assumed to be ignorable given the corresponding treatment and the pretreatment covariates. These are equivalent to assumptions 1, 2, 4, 6, 7, and 8. The estimation involves stratifying a sample on pretreatment covariates. Within each stratum, the average causal mediated effect (i.e., the natural indirect effect) can be estimated by obtaining sample estimates of the conditional outcome associated with a given treatment and a given mediator value, the difference in the density of a given mediator value between the treatment group and the control group, and the product of these two quantities summed over all the mediator values. Finally, the causal effect estimate is obtained by aggregating the stratum-specific estimates.

## 4. Ratio of Mediator Probability Weighting

### 4·1 Statistical Adjustment for Pretreatment Covariates

I propose an alternative non-parametric weighting approach for estimating both natural direct and indirect effects. This new method invokes the same assumptions as those required by the conditional structural model approach described in the previous section. These assumptions suffice for estimating natural direct and indirect effects in the absence of post-treatment covariates.

In general, the joint distribution of the observed data $O = (X, A, Z_A, Y_{AZ_A})$ can be represented as
$$f^{(a,z)}(Y_{az}|A = a, Z_a = z, X) \times q^{(a)}(Z_a = z|A = a, X) \times p(A = a|X) \times h(X),$$
where $f^{(a,z)}(\cdot)$, $q^{(a)}(\cdot)$, $p(\cdot)$, and $h(\cdot)$ are density functions. For simplicity, I use $f(\cdot)$ to represent $f^{(a,z)}(\cdot)$ in the discussion below.

When $A$ is binary, to estimate the natural direct effect $E(Y_{1Z_0} - Y_{0Z_0})$ and the natural indirect effect $E(Y_{1Z_1} - Y_{1Z_0})$, we need an estimate of each of the three marginal mean outcomes $E(Y_{0Z_0})$, $E(Y_{1Z_1})$, and $E(Y_{1Z_0})$. An unbiased estimate of $E(Y_{0Z_0})$ and $E(Y_{1Z_1})$ can each be obtained from the observed data under Assumptions 1 and 8. In its general form,
$$E(Y_{aZ_a}) \equiv E\{E(Y_{aZ_a}|X)\} = E\{E(Y_{aZ_a}|A = a, X)\} = E(Y^*|A = a), \qquad (1)$$

where $Y^* = W_{(aZ_a)}Y$ and $W_{(aZ_a)} = p(A = a)/p(A = a|X)$ for all possible values of $a$. This is equivalent to the inverse-probability-of-treatment weight (IPTW) used in marginal structural models (Robins, 1999; Rosenbaum, 1987). I will discuss in section 4·3 alternative strategies for computing the weight to increase the robustness of the estimation results.

I develop a different type of weighting—namely, ratio of mediator probability weighting—to estimate $E(Y_{1Z_0})$. Under Assumptions 4 and 7, that is, no confounding of mediator-outcome relationship either within a treatment condition or across treatment conditions, we have that
$$f(Y_{az} = y|A = a, Z_{a'} = z, X = x) = f(Y_{az} = y|A = a, Z_a = z, X = x).$$
In other words, within levels of the pretreatment covariates, the counterfactual outcome $Y_{1Z_0}$ of experimental units when their counterfactual mediator $Z_0$ would display value $z$ is assumed to be the same in distribution as their observed outcome $Y_{1Z_1}$ when the observed mediator $Z_1$ actually displays value $z$. Moreover, under Assumption 6, the experimental group and the control group are exchangeable in distribution of $Z_0$ within levels of the pretreatment covariates. Hence, the basic rationale is to take the integral of the observed outcome of the experimental units over the conditional distribution of $Z_0$, $q^{(0)}(Z_0 = z|A = 0, X)$. This is equivalent to assigning a weight to each experimental unit such that the weighted distribution of $Z_1$ in the experimental group approximates the distribution of $Z_0$ in the control group within levels of the pretreatment covariates. The weight to be applied for this purpose is
$$q^{(0)}(Z_0 = z|A = 0, X)/q^{(1)}(Z_1 = z|A = 1, X).$$
In addition, we apply weight $p(A = 1)/p(A = 1|X)$ to make the experimental group and control group exchangeable in pretreatment composition.

THEOREM 1. Under Assumptions 1, 2, 4, 6, 7, and 8, $E(Y^*|A = 1) = E(W_{(1Z_0)}Y|A = 1)$ is an observed data estimand for $E(Y_{1Z_0})$, where $W_{(1Z_0)}$ is equal to
$$\{q^{(0)}(Z_0 = z|A = 0, X)/q^{(1)}(Z_1 = z|A = 1, X)\} \times \{p(A = 1)/p(A = 1|X)\}. \quad (2)$$
In general, to estimate the expectation of the counterfactual outcome $E(Y_{aZ_{a'}})$ in the absence of post-treatment covariates, the weight is
$$W_{(aZ_{a'})} = \{q^{(a')}(Z_{a'} = z|A = a', X)/q^{(a)}(Z_a = z|A = a, X)\} \\ \times \{p(A = a)/p(A = a|X)\}.$$
See Appendix 1 for the proof.

## 4·2 Statistical Adjustment for Both Pretreatment and Post-Treatment Covariates

Let $L_a$ be a vector of post-treatment covariates that confound the mediator-outcome relationship when treatment $a$ is given. In general, the joint distribution of the observed data $O = (X, A, L_A, Z_A, Y_{AZ_A})$ can be represented as
$$f(Y_{az}|A = a, Z_a = z, X, L_a) \times q^{(a)}(Z_a = z|A = a, X, L_a) \times g^{(a)}(L_a = l|A = a, X) \\ \times p(A = a|X) \times h(X).$$

To adjust for the confounding effects of both pretreatment and post-treatment covariates, I modify the Assumptions 2, 4, and 7 for all values of $a$, $a'$, and $z$:

Assumption 2*. (Nonzero probability of mediator value assignment within a treatment). $0 < pr(Z_a = z \mid A, X, L_a) < 1$.

Assumption 4*. (No confounding of mediator-outcome relationship within a treatment condition). $Y_{az} \amalg Z_a \mid A = a, X, L_a$.

Assumption 7*. (No confounding of mediator-outcome relationship across treatment conditions). $Y_{az} \amalg Z_{a'} \mid A = a, X, L_a$.
In addition, I introduce the following identification assumption:

Assumption 9. (Independence of counterfactual mediator and post-treatment covariates across treatment conditions). $Z_{a'} \amalg L_a \mid A = a, X_{(ix)}$.

Here $X_{(ix)}$ represents a set of pretreatment covariates that may or may not overlap with the pretreatment covariates $X$ required for other assumptions stated above. We use $X^+$ to denote the union of sets of pretreatment covariates $X$ and $X_{(ix)}$.

Even though $Z_{a'}$ and $L_a$ are often related by common causes, Assumption 9 can be made plausible in some settings especially under an appropriate research design. In our application example, a child's pretreatment health conditions are presumably among the most important pretreatment covariates that may confound the relationship between the child's counterfactual attention level $Z_0$ at kindergarten entry as a result of growing up in a non-Head Start setting and the child's health condition $L_1$ as a result of attending Head Start. Moreover, if low-income children from the same neighborhood are assigned at random to Head Start programs versus the control condition, a multi-site randomized design effectively adjusts for neighborhood level confounding factors for the relationship between $Z_0$ and $L_1$. After statistical adjustment for children's pretreatment health conditions among other pretreatment covariates in a multi-site randomized experiment, the assumption that a child's counterfactual attention level $Z_0$ does not depend on the child's health condition $L_1$ may hold approximately. In the presence of post-treatment covariates that confound the mediator-outcome relationships, Assumption 9 appears to be weaker than assuming that such post-treatment covariates do not exist.

When $A$ is binary, again we obtain an unbiased estimate of $E(Y_{0Z_0})$ and $E(Y_{1Z_1})$ from the observed data under Assumptions 1 and 8 in a form similar to that shown in Equation (1).

THEOREM 2. Under Assumptions 1, 2*, 4*, 6, 7*, 8, and 9, $E(Y^*|A = 1) = E(W_{(1Z_0)}Y|A = 1)$ is an observed data estimand for $E(Y_{1Z_0})$ in the presence of post-treatment covariates, where $W_{(1Z_0)}$ is equal to
$$\{q^{(0)}(Z_0 = z|A = 0, X^+)/q^{(1)}(Z_1 = z|A = 1, X^+, L_1)\} \times \{p(A = 1)/p(A = 1|X^+)\}.$$
(3)
In general, to estimate the expectation of the counterfactual outcome $E(Y_{aZ_{a'}})$ in the presence of post-treatment covariates, the weight is
$$W_{(aZ_{a'})} = \{q^{(a')}(Z_{a'} = z|A = a', X^+)/q^{(a)}(Z_a = z|A = a, X^+, L_a)\}$$
$$\times \{p(A = a)/p(A = a|X^+)\}.$$
See Appendix 2 for the proof.

The assumptions listed in Theorem 2 are required for estimating the expected value of the counterfactual outcome $E(Y_{1Z_0})$. Specifically, under Assumptions 4* and 7*, we have that
$$f(Y_{az} = y|A = 1, Z_0 = z, X^+ = x, L_1 = l) = f(Y_{az} = y|A = 1, Z_1 = z, X^+ = x, L_1 = l).$$

Under Assumption 9, we have that
$$q^{(0)}(Z_0 = z|A = 1, X^+ = x, L_1 = l) = q^{(0)}(Z_0 = z|A = 1, X^+ = x).$$
The latter is then equal to $q^{(0)}(Z_0 = z|A = 0, X^+ = x)$ under Assumption 6. The ratio of mediator probability weight to be applied in this case is
$$q^{(0)}(Z_0 = z|A = 0, X^+)/q^{(1)}(Z_1 = z|A = 1, X^+, L_1).$$
Under the assumptions stated above, the weighted distribution of $Z_1$ under the experimental condition approximates the distribution of $Z_0$ under the control condition.

## 4·3 Analytic Procedure

If $A$, $Z_A$, $X$, and $L_A$ are all categorical variables, the conditional probabilities in the numerator and those in the denominator of the weight can be empirically determined by the observed proportions of units in the corresponding cells. This process will reveal any violations of Assumptions 1 and 2 or 2* in the observed data; and if so, the analytic sample will be redefined accordingly by excluding units that show zero probability of assignment to a certain treatment or to a certain mediator value under a given treatment.

In general, for a binary treatment, we can analyze a logistic regression to obtain an estimate of the propensity of being assigned to a particular treatment condition $\theta_A(x) = pr(A = a|X = x)$ (Rosenbaum & Rubin, 1983). Similarly, for a binary mediator, logistic regression can be employed to estimate the propensity of having the mediator value under the given treatment condition, $\theta_{Z_0}(x) = pr(Z_0 = z|A = 0, X = x)$, $\theta_{Z_1}(x) = pr(Z_1 = z|A = 1, X = x)$ or $\theta_{Z_1}(x, l) = pr(Z_1 = z|A = 1, X^+ = x, L_1 = l)$. The propensity score of a mediator measured on an ordinal, categorical, or continuous scale can be estimated by employing alternative strategies proposed in the literature (Huang et al, 2005; Joffe & Rosenbaum, 1999; Imai & van Dyke, 2004; Imbens, 2000; Lu, Zanutto, Hornik, & Rosenbaum, 2001; Zanutto, Lu, & Hornik, 2005).

Here I propose a non-parametric approach to estimating the weight by stratifying the sample on each propensity score. Suppose that we divide the sample into five strata on the basis of $\theta_{Z_0}(x)$ denoted by $S_0 = 1, \cdots, 5$. We then divide the same sample into another five strata on the basis of $\theta_{Z_1}(x)$ denoted by $S_1 = 1, \cdots, 5$. In order to estimate $E(Y_{1Z_0})$ when $A$ is randomized, the ratio-of-mediator-probability weight for a sampled unit $i$ is simply $\left(\frac{n_{S_0, Z_0=1}}{n_{S_0}}\right)/\left(\frac{n_{S_1}}{n_{S_1, Z_1=1}}\right)$. Here $n_{S_0}$ is the number of sampled units in the same propensity stratum with unit $i$ when the sample is stratified on $\theta_{Z_0}(x)$; $n_{S_0, Z_0=1}$ is the number of sampled units in the same propensity stratum that displayed the mediator value $Z_0 = 1$; $n_{S_1}$ is the number of sampled units in the same propensity stratum with unit $i$ when the sample is stratified on $\theta_{Z_1}(x)$; $n_{S_1, Z_1=1}$ is the number of sampled units in the same propensity stratum that displayed the mediator value $Z_1 = 1$. Recent results have shown that weighting on the basis of propensity score stratification produces causal effect estimates that are robust to misspecifications of the propensity score model especially in comparison with IPTW estimates (Hong, in press). The non-parametric weighting approach usually provides a better approximation of nonlinear or non-additive relationships between treatment assignment and pretreatment covariates and has a built-in procedure of excluding units that do not have counterfactual information in the observed data.

The estimation involves analyzing a weighted outcome model as a function of the natural direct effect and the natural indirect effect of interest with minimal model-based assumptions. Each natural effect is represented by a single parameter in the outcome model. When the treatment is binary, I reconstruct the data set to include the sampled control units, the sampled experimental units, and a duplicate set of the experimental units. Let $D$ be a dummy indicator that takes value 1 for the duplicate experimental units and 0 otherwise. We assign the weight as follows: $W = W_{(0Z_0)}$ if $A = 0$ and $D = 0$; $W = W_{(1Z_0)}$ if $A = 1$ and $D = 0$; and $W = W_{(1Z_1)}$ if $A = 1$ and $D = 1$. Here $W_{(0Z_0)} = p(A = 0)/p(A = 0|X)$; $W_{(1Z_1)} = p(A = 1)/p(A = 1|X)$; $W_{(1Z_0)}$ is defined in Equation (2) if there are no post-treatment covariates and is defined in Equation (3) with post-treatment covariates. To simultaneously estimate the natural direct and indirect effects on a continuous outcome, we analyze a weighted regression

$$Y = \gamma_0 + A\gamma_1^{(ND)} + AD\gamma_2^{(NI)} + e.$$

In the above model, $\gamma_0$ represents $E(Y_{0Z_0})$; $\gamma_1^{(ND)}$ represents the natural direct effect $E(Y_{1Z_0} - Y_{0Z_0})$; and $\gamma_2^{(NI)}$ represents the natural indirect effect $E(Y_{1Z_1} - Y_{1Z_0})$. For example, to estimate the natural direct and indirect effects of attending a Head Start program on a child's likelihood of repeating kindergarten, we may analyze a weighted generalized linear model with a logit link function. We use estimated robust standard errors to compute respective confidence intervals for $\gamma_1^{(ND)}$ and for $\gamma_2^{(NI)}$. Alternatively, researchers may use bootstrap to obtain an estimate of the standard error. To improve the precision of estimation, the outcome model may include pretreatment covariates that are strong predictors of the outcome.

## 5. Conclusion

Pearl's (2001) formulation of natural direct and indirect effects promises to infuse inspiration for a new class of scientific questions. Allowing the mediator value under each treatment condition to vary naturally among units in the population, the natural direct effect provides a useful summary of the treatment effect on the outcome when the mediator value remains unchanged by the treatment, while the natural indirect effect summarizes the treatment effect on the outcome attributable to the treatment-induced change in the mediator value. However, applications of this new formulation have been rare due to the practical challenges in implementing the existing methods.

How individuals respond to the treatment assignment at the intermediate stage typically reflects their pretreatment characteristics that may also predict their outcome at a fixed level of the mediator value. Hence treatment-mediator interactions are highly plausible. In the meantime, post-treatment covariates may confound the mediator-outcome relationships in some experimental and non-experimental studies of mediation. This is because the mediator as an intermediate outcome of the treatment could have been influenced by many other processes happening in between. Ignoring either treatment-mediator interactions or post-treatment covariates could lead to misleading results in mediation studies.

I have proposed a weighting approach that accommodates treatment-mediator interactions and adjusts for both pretreatment and post-treatment covariates in estimating the marginal mean of each counterfactual outcome. In particular, in order to estimate $E(Y_{1Z_0})$, the weight is computed as a ratio of the conditional probability of a mediator value under the control condition to the conditional probability of the same mediator value under the experimental condition. The weighted distribution of the mediator under the experimental condition approximates their counterfactual mediator distribution under the control condition. After weighting, the natural direct effect and the natural indirect effect of interest are each represented by a single parameter in a non-parametric outcome model and are estimated simultaneously.

Through weighted estimation of counterfactual means, this new approach displays a number of potentially important advantages over the existing methods for estimating natural direct and indirect effects. First of all, because this new approach estimates the natural direct effect without taking an average over the controlled direct effects, it does not require the assumption of no interaction effect of treatment and mediator on the outcome. Secondly, this new method does not require combining multiple parametric models as has been the case in all other existing methods. Thirdly, the non-parametric outcome model avoids model specification errors and applies regardless of the distribution of the outcome or the functional relationship between the outcome and the mediator. Hence unlike most other existing methods, it does not require an explicit linear or nonlinear relationship between the outcome and the mediator. Fourthly, the weighting approach enables adjustment for a large number of covariates without reducing the degrees of freedom in analyzing the outcome model. Fifthly, the non-parametric approach to estimating the weight enhances the robustness of the causal effect estimates even if the parametric model for each mediator is misspecified. Finally, this new weighting approach enables researchers to adjust for post-treatment covariates that confound the mediator-outcome relationships. Other existing methods require the assumption of no such post-treatment confounders, which may limit their applications. Due to these important features, the non-parametric approach through weighting enables researchers to investigate a significantly broader range of scenarios than most of the existing methods can handle. In the absence of post-treatment covariates, the performance of the ratio-of-mediator-weighting method relative to other non-parametric and parametric methods is yet to be assessed through simulations. Future research may extend this approach to studies of multi-valued treatments, multiple mediators, time-varying treatments, time-varying moderators, and time-varying mediators.

## Acknowledgements

This research was supported by a major research grant from the Spencer Foundation and by a William T. Grant Foundation Scholars Award. The views expressed in this article are those of the author and do not necessarily represent the views of the funding agencies. The author owes special thanks to Tyler VanderWeele and Steve Raudenbush for helpful comments on earlier versions of the paper.

## Appendix 1
## Proof of Theorem 1

Theorem 1 requires that we derive a weight $W_{(1Z_0)}$ such that $E(Y_{1Z_0})$ can be consistently estimated by $E(W_{(1Z_0)}Y|A=1)$.

$$E(Y_{1Z_0}) \equiv E\{E(Y_{1Z_0}|X)\}.$$

By Assumption 8, the above is equal to

$$E\{E(Y_{1Z_0}|A = 1, X)\}$$

$$\equiv \iiint_{x,z,y} y \times f(Y_{1z} = y|A = 1, Z_0 = z, X = x)$$
$$\times q^{(0)}(Z_0 = z|A = 1, X = x) \times h(X = x) dy dz dx,$$

which, by Assumptions 4, 6, and 7, is equal to

$$\iiint_{x,z,y} y \times f(Y_{1z} = y|A = 1, Z_1 = z, X = x) \times q^{(0)}(Z_0 = z|A = 0, X = x)$$
$$\times h(X = x) dy dz dx$$

which, by Bayes Theorem and Assumptions 1 and 2, is equal to

$$\iiint_{x,z,y} y \times f(Y_{1z} = y|A = 1, Z_1 = z, X = x) \times q^{(1)}(Z_1 = z|A = 1, X = x)$$

$$\times h(X = x|A = 1) \times \frac{q^{(0)}(Z_0 = z|A = 0, X = x)}{q^{(1)}(Z_1 = z|A = 1, X = x)}$$

$$\times \frac{p(A = 1)}{p(A = 1|X = x)} dy dz dx = E(Y^*|A = 1),$$

where $Y^* = W_{(1Z_0)}Y$ and
$W_{(1Z_0)} = \{q^{(0)}(Z_0 = z|A = 0, X)/q^{(1)}(Z_1 = z|A = 1, X)\} \times \{p(A = 1)/p(A = 1|X)\}$.
This concludes the proof. □

## Appendix 2
## Proof of Theorem 2

Theorem 2 requires that we derive a weight $W_{(1Z_0)}$ such that $E(Y_{1Z_0})$ can be consistently estimated by $E(W_{(1Z_0)}Y|A = 1)$.

$$E(Y_{1Z_0}) \equiv E\{E(Y_{1Z_0}|X^+)\}.$$

By Assumption 8, the above is equal to

$$E\{E(Y_{1Z_0}|A = 1, X^+)\}$$

$$\equiv \iint_{x,l} \iint_{z,y} y \times f(Y_{1z} = y|A = 1, Z_0 = z, X^+ = x, L_1 = l)$$
$$\times q^{(0)}(Z_0 = z|A = 1, X^+ = x, L_1 = l) \times g^{(1)}(L_1 = l|A = 1, X^+ = x)$$
$$\times h(X^+ = x) dy dz dl dx,$$

which, by Assumptions 4* and 7*, is equal to

$$\iint_{x,l} \iint_{z,y} y \times f(Y_{1z} = y|A = 1, Z_1 = z, X^+ = x, L_1 = l)$$
$$\times q^{(0)}(Z_0 = z|A = 1, X^+ = x, L_1 = l) \times g^{(1)}(L_1 = l|A = 1, X^+ = x)$$
$$\times h(X^+ = x) dy dz dl dx,$$

which, by Assumptions 9 and 6, is equal to

$$\iint_{x,l} \iint_{z,y} y \times f(Y_{1z} = y|A = 1, Z_1 = z, X^+ = x, L_1 = l) \times q^{(0)}(Z_0 = z|A = 0, X^+ = x)$$
$$\times g^{(1)}(L_1 = l|A = 1, X^+ = x) \times h(X^+ = x) dy dz dl dx,$$

which, by Bayes Theorem and Assumptions 1 and 2*, is equal to

$$\iint_{x,l} \iint_{z,y} y \times f(Y_{1z} = y | A = 1, Z_1 = z, X^+ = x, L_1 = l)$$
$$\times q^{(1)}(Z_1 = z | A = 1, X^+ = x, L_1 = l) \times g^{(1)}(L_1 = l | A = 1, X^+ = x)$$
$$\times h(X^+ = x | A = 1) \times \frac{q^{(0)}(Z_0 = z | A = 0, X^+ = x)}{q^{(1)}(Z_1 = z | A = 1, X^+ = x, L_1 = l)}$$
$$\times \frac{p(A = 1)}{p(A = 1 | X^+ = x)} dy dz dl dx$$

$= E(Y^* | A = 1)$,
where $Y^* = W_{(1Z_0)} Y$ and
$$W_{(1Z_0)} = \{q^{(0)}(Z_0 = z | A = 0, X^+) / q^{(1)}(Z_1 = z | A = 1, X^+, L_1)\}$$
$$\times \{p(A = 1) / p(A = 1 | X^+)\}.$$

This concludes the proof. □

## References

Baron, R. M., & Kenny, D. A. (1986). The moderator-mediator variable distinction in social psychological research: Conceptual, strategic, and statistical considerations. *Journal of Personality and Social Psychology*, **51**, 1173-1182.

Duncan, O. D. (1966). "Path analysis: Sociological examples," *American Journal of Sociology*, **72**, 1-16.

Efron, B. (1988). Bootstrap confidence intervals: Good or bad. *Psychological Bulletin*, **104**(2), 293-296.

Holland, P. W. (1986). Statistics and causal inference. *Journal of the American Statistical Association,* **81**, 945-960.

Holland, P. (1988). "Causal inference, path analysis, and recursive structural equations models," *Sociological methodology*, **18**, 449-484.

Hong, G. (In press). Marginal mean weighting through stratification: Adjustment for selection bias in multi-level data. *Journal of Educational and Behavioral Statistics*.

Huang, I-C., Frangakis, C., Dominici, F., Diette, G. B., and Wu, A. W. (2005). Approach for risk adjustment in profiling multiple physician groups on asthma care. *Health Services Research,* **40**, 253-278.

Joffe, M. M., & Rosenbaum, P. R. (1999). Invited commentary: Propensity scores. *American Journal of Epidemiology,* **150**(4), 327-333.

Imai, K., Keele, L., & Yamamoto, T. (2010). Identification, inference and sensitivity analysis for causal mediation effects. *Statistical Science*, **25**(1), 51-71.

Imai, K., & van Dyke, D. A. (2004). Causal inference with general treatment regimes: Generalizing the propensity score. *Journal of the American Statistical Association,* **99**, 854-866.

Imbens, G. (2000). The role of the propensity score in estimating dose-response functions. *Biometrika,* **87**, 706-710.

Lu, B., Zanutto, E., Hornik, R., & Rosenbaum, P. R. (2001). Matching with doses in an observational study of a media campaign against drug abuse. *Journal of the American Statistical Association,* **96**, 1245-1253.

Pearl, J. (2001). Direct and indirect effects. In *Proceedings of the American Statistical Association Joint Statistical Meetings*. Minn, MN: MIRA Digital Publishing, 1572-1581, August 2005.

Pearl, J. (2010). The mediation formula: A guide to the assessment of causal pathways in non-linear models. Los Angeles, CA: University of California, Los Angeles. Technical report R-363, July 2010.

Peterson, M. L., Sinisi, S. E., & van der Laan, M. J. (2006). Estimation of direct causal effects. *Epidemiology,* **17**(3), 276-284.

Robins, J. M. (1999). Marginal structural models versus structural nested models as tools for causal inference. In M. Elizabeth Halloran and Donald Berry (Eds.), *Statistical Models in Epidemiology, the Environment, and Clinical Trials* (pp.95-134). New York: Springer.

Robins, J. M. (2003). Semantics of causal DAG maodels and the identification of direct and indirect effects. In *Highly Structural Stochastic Systems* (P.J. Green, N. L., Hjort and S. Richardson, eds.) 70-81. Oxford: Oxford University Press.

Robins, J. M. & Greenland, S. (1992). Identifiability and exchangeability for direct and indirect effects. *Epidemiology,* **3**(2), 143-155.

Rosenbaum, P. R. (1984). The consequences of adjustment for a concomitant variable that has been affected by the treatment. *Journal of the Royal Statistical Society*, *Series A*, **147**, 656-666.

Rosenbaum, P. R. (1987). Model-based direct adjustment. *Journal of the American Statistical Association*, **82**, 387-394.

Rosenbaum, P. R., and Rubin, D. B. (1983), The central role of the propensity score in observational studies for causal effects, *Biometrika*, **70**, 41-55.

Rubin, D. B. (1978), Bayesian inference for causal effects: The role of randomization. *The Annals of Statistics*, **6**, 34-58.

Rubin, D. B. (1986), "Comment: Which Ifs Have Causal Answers." *Journal of the American Statistical Association*, **81**, 961-962.

Sobel, M. E. (2008). Identification of causal parameters in randomized studies with mediating variables. *Journal of Educational and Behavioral Statistics*, **33**(2), 230-251.

van der Lann, M. J., & Peterson, M. L. (2008). Direct effect models. *The International Journal of Biostatistics,* **4**(1), Article 23.

VanderWeele, T. (2009). Marginal structural models for the estimation of direct and indirect effects. *Epidemiology,* **20**, 18-26.

VanderWeele, T., & Vansteelandt, S. (2009). Conceptual issues concerning mediation, interventions, and composition. *Statistics and its Interface,* **2**, 457-468.

Zanutto, E., Lu, Bo., & Hornik, R. (2005). Using propensity score subclassification for multiple treatment doses to evaluate a national antidrug media campaign. *Journal of Educational and Behavioral Statistics,* **30**(1), 59-73.